\documentstyle[12pt]{article}
\begin{document}
\setlength{\baselineskip}{0.30in}
\newcommand{\be}{\begin{eqnarray}}
\newcommand{\ee}{\end{eqnarray}}
\newcommand{\bi}{\bibitem}

\begin{center}
{\Large \bf {Mystery of Vacuum Energy \\
or  \\
Rise and Fall of Cosmological Constant
 }
}
\bigskip
\\
{\bf A.D. Dolgov }
 \\[.05in]
{\it{
INFN,
Via Paradiso, 12 - 44100 Ferrara Italy\\
and \\
{ ITEP, Bol. Cheremushkinskaya 25, Moscow 113259, Russia.}
}}

\end{center}
\begin{abstract}

Two sides of cosmological constant problem are discussed: a mysterious 
compensation of all contributions to vacuum energy with the accuracy of
100-50 orders of magnitude and a surprising equality of a constant vacuum
energy density to the present-day value of time dependent cosmological
energy density.

\end{abstract}
 
Cosmological constant was born in 1918 in Einstein's paper~\cite{ae},
where he noticed that the equations of General Relativity would possess
a stationary cosmological solution if one added an extra term proportional
to metric tensor $g_{\mu\nu}$ with a {\it constant} coefficient $\Lambda$:
\be
R_{\mu\nu} -{1\over 2} g_{\mu\nu} R = 8\pi G_N T_{\mu\nu} +
g_{\mu\nu} \Lambda
\label{eineq}
\ee
For quite a long time the parameter $\Lambda$ was considered as an extra degree 
of freedom in GR, permitted by general covariance. Now it is commonly 
understood that $\Lambda$ should be identified with vacuum energy density:
\be
\rho_{vac} = \Lambda /8\pi G_N \equiv \Lambda m_{Pl}^2 /8\pi
\label{rhovac}
\ee
Naively one might think that vacuum is something which is empty and thus its
energy must be zero. That is why many physicists and especially astronomers 
were (and still are) very 
strongly against cosmological constant. Einstein himself considered 
$\Lambda$ as the biggest blunder of his life. The point of view of one of 
the creators of the big bang cosmology, George Gamow, is expressed by his
words~\cite{gg}: "$\lambda$ again raises its nasty head" (he used a small 
letter, which somewhat reminded a snake, possibly to show his contempt). The
attitude of astro-cosmo-physical society toward $\Lambda$ was (and
possibly is) strongly polarized. The majority did not accept the notion that
vacuum might gravitate and insisted that $\Lambda \equiv 0$. On the other
hand, some non-negligible by number and quality scientists believed that 
$\Lambda$ exists and might be cosmologically important. Among them are such
distinguished names as Lema\^itre, De Sitter, and Eddington. By 
Lema\^itre's opinion, even if Einstein did not do anything except 
discovering $\Lambda$-term it would be enough to make his name famous.

A striking fine-tuning, however, justified a negative attitude 
to a cosmologically essential $\Lambda$-term. Vacuum energy must remain
constant in the course of the universe expansion, while the energy density 
of usual matter decreases roughly as critical energy density
$\rho_c\sim m^2_{Pl} /t^2$. Thus, if {\it now} $ \rho_{vac} $ is close 
to $\rho_c$, it looks as a very strange coincidence. This problem stimulated 
the notion of the so called time dependent cosmological "constant". This idea 
was first put forward in 1932 by M. Bronstein~\cite{mb} 
However this assumption is not as 
innocent as it sounds, and Bronstein's conjecture was justly criticized by 
Landau, as is admitted in Bronstein's paper~\cite{mb}. Similar criticism is 
valid for all subsequent development of the idea; the list of publication 
on the subject is quite long and can be, at least partly, found in 
refs.~\cite{ad1}-\cite{more-rev}. 
In brief, the problem with $\Lambda =\Lambda (t)$ is the 
following. Taking covariant derivative of both sides of eq.~(\ref{eineq}),
one finds:
\be
\partial_\mu \Lambda + 8\pi G_N T_{\mu;\nu}^{\nu}  =0
\label{dlam}
\ee
Usually the energy-momentum tensor of matter is obtained from the 
matter action by variation over metric:
\be 
T_{\mu\nu} = {\delta \over \delta g^{\mu\nu} }
\int d^4 x \sqrt {-g} \,{\cal L}
\label{tmunu}
\ee
and if general covariance is unbroken, $T_{\mu\nu}$ is conserved as a 
result of equations of motion, $T_{\mu;\nu}^{\nu}  =0$ and
$\partial_\mu \Lambda =0$. Thus a high price is to be paid for an 
introduction
a time-dependent $\Lambda$-term: either one has to reject the standard 
Lagrangian formalism or to invent a new dynamical field which is a highly
non-trivial task.

There is one more and even much more striking side of the cosmological 
constant problem. Even if vacuum energy is cosmologically significant, i.e.
$\rho_{vac} \sim \rho_c$, it is unreasonably tiny in terms of particle 
physics scale:
\be
 \rho_{vac} \leq \rho_c \sim 10^{-47} {\rm GeV}^4
\label{rhovac2}
\ee
Moreover there are contributions into vacuum energy, which are  50-100 
orders of magnitude(!) larger than this upper limit. The first review
papers stressing the importance of the vacuum energy problem were
published in 1989~\cite{sw1,ad2}. Since then several more reviews appeared
~\cite{more-rev} where one can find more up-to-date references.
 
There are vacuum fluctuations which naively 
have infinitely large energy density. Fortunately in the world with
equal number of bosonic and fermionic species this infinity cancels out,
as was noticed by Zeldovich~\cite{yabz} a few years before the pioneering
papers on supersymmetry were published~\cite{susy}. Still since supersymmetry
is not exact, only infinities are compensated but finite non-compensated 
remnants are of the order of the SUSY breaking scale,
$\rho_{vac}^{(susy)} \sim m^4_{susy} \geq 10^8\,\, {\rm GeV}^4$.
Possibly there were phase transitions in the early universe
in the course of which vacuum energy changed by $10^{60}\,\, {\rm GeV}^4$, if 
they took place at GUT scale or by $10^{10} \,\,{\rm GeV}^4$ at electroweak
scale. One could argue however that 
these phase transitions are manifestations of high energy
physics and who knows, if they existed or not. 
Still there exist some other contributions which, though
smaller than the grand unification and even the electroweak ones, are 
enormous in comparison with $10^{-47} \,{\rm GeV}^4$. It is well known that
vacuum state in quantum chromodynamics (QCD) is not empty. It is filled by
quark (or chiral)~\cite{gor} and gluon~\cite{svz} condensates.
The existence of these condensates is practically an 
experimental fact. Successful QCD description of hadron properties is 
impossible without them. The vacuum energy density of the quark
condensate, $\langle \bar q q \rangle$, is about $10^{-4}\,{\rm GeV}^4$ and 
that of gluon condensate, $\langle G_{\mu\nu}^2 \rangle$, is approximately an 
order of magnitude bigger. Comparing these numbers with the upper 
bound~(\ref{rhovac2}) we see that there {\it must} exist something 
which does
not know anything about quarks and gluons (this "something" is not related
to quarks and gluons by  the usual QCD interactions, otherwise it will be
observed in experiment) but still this mysterious agent is able to 
compensate vacuum energies of quarks and gluons with the fantastic 
accuracy of $10^{-44}$.

Several possibilities to solve this mystery were discussed in the literature,
none was successful. One can imagine logically the following four ways
(however, it is quite possible that a number 5 is realized):\\
1. Modification of gravity on macroscopic distances (possibly due to
higher dimensions).\\
2. Anthropic principle.\\
3. A symmetry leading to $\rho_{vac} = 0$.\\
4. Adjustment mechanism (either by a new massless or very light field
or due to infrared instability of quantum fluctuations in De Sitter
space-time).\\

To modify gravity at big distances, so that the 
vacuum part of energy-momentum 
tensor does not gravitate, is a formidable task, keeping in mind that 
general covariance, which implies, in particular, covariant conservation of 
energy momentum tensor and ensures vanishing of the graviton mass, 
must be respected. In the course of the universe evolution equation of 
state may change so that vacuum energy changes (it usually happens in
phase transitions) and subtraction of vacuum energy seems incompatible with
conservation of total $T_{\mu\nu}$.

Anthropic principle is possibly the last resort in the case that
no other solution can be found. At the moment the situation 
with $\Lambda$-term
reminds the one that existed in Friedman cosmology before inflationary 
resolution of seemingly unsolvable cosmological problems has been 
proposed \cite{guth}.

It would be very attractive to discover  
a symmetry principle forbidding vacuum energy.
Such a symmetry should connect known
fields with new unknown ones. Some of those fields should be very light
to achieve the cancellation on the scale $10^{-3}$ eV. Neither such fields 
are observed, nor such a symmetry is known. Moreover, if cosmological constant
is not precisely zero, as strongly indicated by the recent data~\cite{sn1}, 
one has to explain why a symmetry breaking produces a remnant very 
close to $\rho_c$ today. 

For me the adjustment mechanism seems to be the most promising at 
the present time, though this point of view is possibly not shared by
many physicists working on the problem. The idea of adjustment
is similar to the solution of CP-conservation problem  
in quantum chromodynamics by the axion field~\cite{pq}-\cite{wil}.
Assume that there exists a new massless field $\Phi$
coupled to gravity in such a way 
that this field is unstable in De Sitter background. Such a field would
develop vacuum condensate whose energy could kill the source~\cite{ad3}.
At the moment no satisfactory adjustment mechanism, which gives a realistic
cosmology, has been found. However, independently of concrete realization,
adjustment models generically predict that vacuum energy is compensated 
only up to the terms of the order of 
$\rho_c (t)$~\cite{ad3,ad1}. The non-compensated remnants may possess a 
rather peculiar equation of state. 

A natural idea that $\Phi$ is a scalar field meets 
serious difficulties~\cite{sw1} but higher spin fields may be more 
perspective. In particular, a second rank symmetric tensor $S_{\mu\nu}$
with a very simple Lagrangian density:
\be 
{\cal L} =   S_{\mu\nu;\alpha}S^{\mu\nu;\alpha} 
\label{smunu}
\ee
could develop a condensate of isotropic components $S_{tt}$ 
and $S_{ij} \sim \delta_{ij}$ which eliminate an original $\rho_{vac}$ down to
terms of the order $m_{Pl}^2/t^2$ and successfully change De Sitter expansion 
into a power law one. 
Still the model does not lead to realistic cosmology. In particular, 
the concrete realization proposed in ref.~\cite{ad1} leads to a strong
time variation of the gravitational constant~\cite{rubakov99}.
There are quite many different models of adjustment mechanism discussed
in the literature. The references can be found in the above cited
reviews~\cite{more-rev}; recently there appeared a few more
papers~\cite{rubakov00ad}. 

Another model of adjustment discussed in the literature~\cite{infrared}
is based on the known phenomenon that quantum fluctuations of massless
fields in De Sitter space-time are infrared unstable. The effects of such
instability are significant but it is not yet clear if they could kill
their creator, vacuum energy.

An interesting approach to self-tuning of
the cosmological constant in multidimensional theories
is actively developed during last years. The number of papers on the subject
is already impressive and constantly rising. An incomplete list
of references includes the papers~\cite{multid} which may help to 
understand some ideas showing how this mechanism may operate. One
should keep in mind however, that we live in four dimensional world
(higher dimensions, if they exist, are either small or penetration there is
protected by a potential wall) and all the
mechanisms of nullification of vacuum energy in higher dimensions must
be described in 4D language as well. If new higher dimensional
physics starts to operate at TeV scale it is not clear how it may 
cancel out QCD vacuum condensates of quarks and gluons which are 
developed at 100 MeV scale. 

One may try another less ambitious approach - to neglect the problem of
extremely large vacuum energy introduced by particle physics and quantum
field theory and to try simply to describe cosmology phenomenologically 
introducing an additional parameter, or better to say, a function 
$\Lambda (t)$, keeping in mind, of course, that this can be done by
introducing a new light or massless field to keep general covariance unbroken. 
Earlier papers on the subject include refs.~\cite{declam}.
This approach attracted great attention after indications for accelerated
cosmological expansion in 1998~\cite{sn1}. An unknown form of energy was
mimicked by a scalar field with the equation of state $p=w\rho$ with
a negative parameter $w< -1/3$. The name ``quintessence'' was suggested
for this field~\cite{quint}. This work stimulated a lot of activity in the
field. Many references can be found in the reviews~\cite{more-rev}. My 
special pleasure is to cite in this connection the works by M.~Novello 
and his collaborators~\cite{novello}.

A question of vital importance for all the models that are aimed to a 
resolution of vacuum energy problem is what is the magnitude of 
vacuum or vacuum-like energy (now the term ``dark energy'' is 
commonly used for the latter). If it is too large by absolute value,
cosmology would be very much different and our type of life would be
hardly possible~\cite{anthr} (anthropic principle?).
If vacuum energy is too small it would not be observed. According to the
present day data, $\Omega_{vac} \approx 0.7$. Thus we are very lucky,
it is large enough to be observed in cosmological phenomena and 
small enough not to spoil our life. Moreover, as we have already mentioned
above, this adds another interesting mystery to the list of unexplained
phenomena: why $\rho_{vac}$ which stays constant during cosmological
evolution is so close to energy density of matter which evolves with
cosmic time as $\rho_m \sim 1/t^2$. There is indeed weird cosmic conspiracy:
all forms of matter have similar contributions to cosmological energy
density: usual baryonic matter ($\Omega_b \approx 0.05$), 
dark matter ($\Omega_{DM} \approx 0.2-0.3$) - possibly dark matter consists
of several components of comparable mass/energy density, and now there
is another member of the club, vacuum energy with $\Omega_{vac} \approx 0.7$. 

Though there are quite strong data now
indicating that seemingly empty space indeed (anti)gravitates,
still this conclusion is so revolutionary that it 
deserves more thorough checks.
During practically all previous (XXth) century the overwhelming feeling
was that cosmological constant is identically zero. There were only 2-3
short periods when a non-zero lambda was seriously considered. After the
Einstein's ``blunder'' of 1918 the most serious was one at the end of
60's when astronomical data indicated an accumulation of quasars near
the red-shift $z=2$~\cite{petrosian67}. Then again cosmological constant
was strongly out of fashion till the end of the century when 
different pieces of data have accumulated, all implying that $\Lambda$
is most probably non-vanishing. Prior to the observation of large z
supernovae~\cite{sn1}, there were indications to the age crisis. The
universe age can be expressed through the present day value of the Hubble
parameter $H = 100 h$ km/sec/Mpc and the cosmological energy density as:
\be
t_u = {9.788\cdot 10^9\,h^{-1} {\rm yr} } 
\int_0^1 {dx \over \sqrt{ 1 - \Omega_{tot} +
\Omega_{m} x^{-1} + \Omega_{rel} x^{-2} + \Omega_{vac}x^2 }}
\label{tu}
\ee
where $\Omega_{m}$, $\Omega_{rel}$, and $\Omega_{vac}$
correspond respectively to the energy density of nonrelativistic
matter, relativistic matter, and to the vacuum energy density
(or, what is the same, to the cosmological constant);
$\Omega_{tot} = \Omega_{m} + \Omega_{rel} + \Omega_{vac}$.
If $\Omega_{vac} = 0$, $\Omega_{tot} =1$ (according to inflation), 
and $h \approx 0.7$ (see e.g. ref.~\cite{freedman01})
then $t_u$ would be 9.3 Gyr, much smaller than the age of the
universe estimated from nuclear chronology and the ages of old globular 
clusters, for a recent analysis see e.g. ref.~\cite{age}. However if 
$\Omega_{vac} > 0$ the age crisis disappears. In particular, for
$\Omega_{tot} =1$ and $\Omega_{vac} = 0.7$ the universe age would be
13.5 Gyr in good agreement with the age determined by other methods.

Observations of high red-shift supernovae, if they are standard candles, 
allow to measure deceleration parameter, $q = \Omega_m -2\Omega_{vac}$.
The two contributions enter with opposite sign because positive vacuum
energy antigravitates. Indeed, according to the Einstein equations 
effective source of gravitational force is $\rho +3p = -2\rho$, where 
$\rho$ and $p$ are respectively energy and pressure densities and for
vacuum $p=-\rho$.
On the other hand, position of the first acoustic peak in the angular
spectrum of cosmic microwave background radiation (CMBR) permits to 
conclude that $\Omega_{tot} = \Omega_m + \Omega_{vac} =1$. One can
also determine $\Omega_m$ directly measuring peculiar velocities,
gravitational lensing, or evolution of cluster abundances. All these
methods give $\Omega_m \approx 0.3$. 
The determination of three different combinations of $\Omega_m$ and 
$\Omega_{vac}$ mentioned above are analyzed e.g. in ref.~\cite{bridge00}
(there are many more works where a similar conclusion is reached but their
list is too long for this brief paper).
It is argued that all three data sets well agree converging to 
$\Omega_{vac}\approx 0.7$ and $\Omega_m \approx 0.3$.
So it seems that now and at last astronomy presents a solid piece of evidence
in favor of non-zero and significant cosmological term.

To summarize, the problem of vacuum energy remains possibly the most profound
problem of contemporary physics. It is a unique example when theoretical 
expectation differ from observation by 100-50 orders of magnitude.
The recent indications that the universe expands with acceleration and that
cosmological constant (or an unknown form of dark energy) is 
non-vanishing significantly amplified gravity of the problem. 
If earlier one might think 
that vacuum does not gravitate at all, now it seems that empty space creates
gravitational repulsion. This observation improved the status of 
adjustment mechanism which predicted that there must be some unusual form
of energy due to a non-complete compensation of vacuum~\cite{ad3,ad1}
energy. Still no satisfactory form of adjustment mechanism leading to
realistic cosmology has been yet found. Hopefully when it is achieved the 
theory will indicate a unique and well defined form of dark energy but at
the moment the problem is far from resolution. It presents a strong 
challenge for fundamental research to and a serious indication to
new physics. Two famous small clouds on the fundamental physics sky
a century ago brought to life revolutionary quantum mechanics and relativity 
theory.  Quite possibly the mystery of vacuum energy will stimulate new 
ideas in physics of this century (or millennium?).


\end{document}